\documentclass{scrartcl}

\usepackage[a4paper,textwidth=170mm,textheight=240mm]{geometry}

\usepackage[T1]{fontenc}
\usepackage{microtype}
\usepackage[inline]{enumitem}
\usepackage{siunitx}
\usepackage{graphicx}
\usepackage[table]{xcolor}

\usepackage{nicematrix}

\usepackage[super,sort&compress]{natbib}
\usepackage{url}
\usepackage{booktabs}
\usepackage{pifont}

\usepackage{tikz}
\usepackage{pgfplots}
\usepackage{pgfplotstable}
\pgfplotsset{
    compat=1.18,
    axis x line*=bottom,
    axis y line*=left,
    discard if not/.style 2 args={
        filter discard warning=false,
        x filter/.code={
            \edef\tempa{\thisrow{#1}}
            \edef\tempb{#2}
            \ifx\tempa\tempb%
            \else
            
            \fi
        }
    }
}
\usepgfplotslibrary{groupplots}
\usetikzlibrary{arrows.meta,fit,positioning}
\definecolor{rwth-blue-100}{cmyk/RGB/HTML/gray}{1,.50,0,0/0,84,159/00549F/1}
\definecolor{rwth-blue-75}{cmyk/RGB/HTML/gray}{.75,.38,0,0/64,127,183/407FB7/.75}%
\definecolor{rwth-blue-50}{cmyk/RGB/HTML/gray}{.45,.14,0,0/142,186,229/8EBAE5/.5}
\definecolor{rwth-blue-25}{cmyk/RGB/HTML/gray}{.23,.07,0,0/199,221,242/C7DDF2/.25}%
\definecolor{rwth-blue-10}{cmyk/RGB/HTML/gray}{.09,.03,0,0/232,241,250/E8F1FA/.1}%
\definecolor{rwth-black-100}{cmyk/RGB/HTML}{0,0,0,1/0,0,0/000000}%
\definecolor{rwth-black-75}{cmyk/RGB/HTML}{0,0,0,.75/100,101,103/646567}%
\definecolor{rwth-black-50}{cmyk/RGB/HTML}{0,0,0,.5/156,158,159/9C9E9F}%
\definecolor{rwth-black-25}{cmyk/RGB/HTML}{0,0,0,.25/207,209,210/CFD1D2}%
\definecolor{rwth-black-10}{cmyk/RGB/HTML}{0,0,0,.1/236,237,237/ECEDED}%
\definecolor{rwth-magenta-100}{cmyk/RGB/HTML}{0,1,.25,0/227,0,102/E30066}%
\definecolor{rwth-magenta-75}{cmyk/RGB/HTML}{0,.75,.19,0/233,96,136/E96088}%
\definecolor{rwth-magenta-50}{cmyk/RGB/HTML}{0,.5,.13,0/241,158,177/F19EB1}%
\definecolor{rwth-magenta-25}{cmyk/RGB/HTML}{0,.25,.06,0/249,210,218/F9D2DA}%
\definecolor{rwth-magenta-10}{cmyk/RGB/HTML}{0,.10,.03,.02/253,238,240/FDEEF0}%
\definecolor{rwth-yellow-100}{cmyk/RGB/HTML}{0,0,1,0/255,237,0/FFED00}%
\definecolor{rwth-yellow-75}{cmyk/RGB/HTML}{0,0,.75,0/255,240,85/FFF055}%
\definecolor{rwth-yellow-50}{cmyk/RGB/HTML}{0,0,.50,0/255,245,155/FFF59B}%
\definecolor{rwth-yellow-25}{cmyk/RGB/HTML}{0,0,.25,0/255,250,209/FFFAD1}%
\definecolor{rwth-yellow-10}{cmyk/RGB/HTML}{0,0,.10,0/255,253,238/FFFDEE}%
\definecolor{rwth-petrol-100}{cmyk/RGB/HTML}{1,.3,.5,.3/0,97,101/006165}%
\definecolor{rwth-petrol-75}{cmyk/RGB/HTML}{.75,.23,.38,.23/45,127,131/2D7F83}%
\definecolor{rwth-petrol-50}{cmyk/RGB/HTML}{.5,.15,.25,.15/125,164,167/7DA4A7}%
\definecolor{rwth-petrol-25}{cmyk/RGB/HTML}{.25,.08,.13,.08/191,208,209/BFD0D1}%
\definecolor{rwth-petrol-10}{cmyk/RGB/HTML}{.1,.03,.05,.03/230,236,236/E6ECEC}%
\definecolor{rwth-turquoise-100}{cmyk/RGB/HTML}{1,0,.4,0/0,152,161/0098A1}%
\definecolor{rwth-turquoise-75}{cmyk/RGB/HTML}{.75,0,.3,.0/0,177,183/00B1B7}%
\definecolor{rwth-turquoise-50}{cmyk/RGB/HTML}{.5,0,.2,0/137,204,207/89CCCF}%
\definecolor{rwth-turquoise-25}{cmyk/RGB/HTML}{.25,0,.1,0/202,231,231/CAE7E7}%
\definecolor{rwth-turquoise-10}{cmyk/RGB/HTML}{.1,0,.04,0/235,246,246/EBF6F6}%
\definecolor{rwth-green-100}{cmyk/RGB/HTML}{.7,0,1,0/87,171,39/57AB27}%
\definecolor{rwth-green-75}{cmyk/RGB/HTML}{.52,0,.75,0/141,192,96/8DC060}%
\definecolor{rwth-green-50}{cmyk/RGB/HTML}{.35,0,.5,0/184,214,152/B8D698}%
\definecolor{rwth-green-25}{cmyk/RGB/HTML}{.18,0,.25,0/221,235,206/DDEBCE}%
\definecolor{rwth-green-10}{cmyk/RGB/HTML}{.07,0,.1,0/242,247,236/F2F7EC}%
\definecolor{rwth-maygreen-100}{cmyk/RGB/HTML}{.35,0,1,0/189,205,0/BDCD00}%
\definecolor{rwth-maygreen-75}{cmyk/RGB/HTML}{.26,0,.75,0/208,217,92/D0D95C}%
\definecolor{rwth-maygreen-50}{cmyk/RGB/HTML}{.18,0,.5,0/224,230,154/E0E69A}%
\definecolor{rwth-maygreen-25}{cmyk/RGB/HTML}{.09,0,.25,0/240,243,208/F0F3D0}%
\definecolor{rwth-maygreen-10}{cmyk/RGB/HTML}{.04,0,.1,0/249,250,237/F9FAED}%
\definecolor{rwth-orange-100}{cmyk/RGB/HTML}{0,.4,1,0/246,168,0/F6A800}%
\definecolor{rwth-orange-75}{cmyk/RGB/HTML}{0,.3,.75,0/250,190,80/FABE50}%
\definecolor{rwth-orange-50}{cmyk/RGB/HTML}{0,.2,.5,0/253,212,143/FDD48F}%
\definecolor{rwth-orange-25}{cmyk/RGB/HTML}{0,.1,.25,0/254,234,201/FEEAC9}%
\definecolor{rwth-orange-10}{cmyk/RGB/HTML}{0,.04,.1,0/255,247,234/FFF7EA}%
\definecolor{rwth-red-100}{cmyk/RGB/HTML}{.15,1,1,0/204,7,30/CC071E}%
\definecolor{rwth-red-75}{cmyk/RGB/HTML}{.11,.75,.75,0/216,92,65/D85C41}%
\definecolor{rwth-red-50}{cmyk/RGB/HTML}{0,.35,.47,.1/230,150,121/E69679}%
\definecolor{rwth-red-25}{cmyk/RGB/HTML}{0,.16,.23,.05/243,205,187/F3CDBB}%
\definecolor{rwth-red-10}{cmyk/RGB/HTML}{0,.06,.09,.02/250,235,227/FAEBE3}%
\definecolor{rwth-bordeaux-100}{cmyk/RGB/HTML}{.25,1,.70,.20/161,16,53/A11035}%
\definecolor{rwth-bordeaux-75}{cmyk/RGB/HTML}{.19,.75,.52,15/182,82,86/B65256}%
\definecolor{rwth-bordeaux-50}{cmyk/RGB/HTML}{.13,.5,.35,.1/205,139,135/CD8B87}%
\definecolor{rwth-bordeaux-25}{cmyk/RGB/HTML}{.06,.25,.18,.05/229,197,192/E5C5C0}%
\definecolor{rwth-bordeaux-10}{cmyk/RGB/HTML}{.03,.1,.07,.02/245,232,229/F5E8E5}%
\definecolor{rwth-violet-100}{cmyk/RGB/HTML}{.70,1,.35,.15/97,33,88/612158}%
\definecolor{rwth-violet-75}{cmyk/RGB/HTML}{.52,.75,.26,.11/131,78,117/834E75}%
\definecolor{rwth-violet-50}{cmyk/RGB/HTML}{.35,.5,.18,.08/168,133,158/A8859E}%
\definecolor{rwth-violet-25}{cmyk/RGB/HTML}{.18,.25,.09,.04/210,192,205/D2C0CD}%
\definecolor{rwth-violet-10}{cmyk/RGB/HTML}{.07,.1,.04,.02/237,229,234/EDE5EA}%
\definecolor{rwth-purple-100}{cmyk/RGB/HTML}{.6,.6,0,0/122,111,172/7A6FAC}%
\definecolor{rwth-purple-75}{cmyk/RGB/HTML}{.45,.45,0,0/155,145,193/9B91C1}%
\definecolor{rwth-purple-50}{cmyk/RGB/HTML}{.3,.3,0,0/188,181,215/BCB5D7}%
\definecolor{rwth-purple-25}{cmyk/RGB/HTML}{.15,.15,0,0/222,218,235/DEDAEB}%
\definecolor{rwth-purple-10}{cmyk/RGB/HTML}{.06,.06,0,0/242,240,247/F2F0F7}%

\pgfplotscreateplotcyclelist{rwth}{
    {fill=rwth-blue-100,draw=rwth-blue-100!80!black},
    {fill=rwth-maygreen-100,draw=rwth-maygreen-100!80!black},
    {fill=rwth-orange-100,draw=rwth-orange-100!80!black},
    {fill=rwth-petrol-100,draw=rwth-petrol-100!80!black},
    {fill=rwth-violet-100,draw=rwth-violet-100!80!black},
    {fill=rwth-red-100,draw=rwth-red-100!80!black},
    {fill=rwth-purple-100,draw=rwth-purple-100!80!black}
}

\title{A repository for discovery and reuse of higher-order network datasets}
\author{%
    Florian Frantzen$^{1,*}$ and Michael T. Schaub$^{1}$\\
    $^{1}$Faculty of Computer Science, RWTH Aachen University, Aachen, Germany\\
    $^{*}$Correspondence: \texttt{frantzen@netsci.rwth-aachen.de}
}
\date{}

\begin{document}

\maketitle

\begin{abstract}
    Higher-order network datasets are dispersed across publications, institutional archives, and software-specific collections, making them difficult to discover, compare, and reuse.
    We introduce the Aachen Higher-Order Repository of Networks (\texttt{AHORN}), a curated repository of standardized higher-order network datasets derived from publicly released sources.
    Each dataset entry links a converted dataset to its source, metadata, citation guidance, conversion code, and version history.
    The repository supports browsable and machine-readable discovery, revision-specific downloads, format validation, and exports for interoperable reuse.
    We describe the repository architecture, curation workflow, access tools, and the coverage and limitations of the catalog snapshot analyzed in this article.
\end{abstract}

\section{Introduction}%
\label{section:introduction}

Many complex systems contain interactions involving more than two entities \citep{Battiston:2020,Battiston:2021}.
Examples include coauthorship groups \citep{Benson:2018}, protein interaction systems \citep{Murgas:2022}, and social contacts \citep{Neuhäuser:2022}.
Representing these systems only as dyadic graphs can obscure structural and dynamical features created by multi-way relationships \citep{Battiston:2020,Battiston:2021,Benson:2018,Murgas:2022,Neuhäuser:2022}.

Accordingly, higher-order network formalisms, including hypergraphs and simplicial complexes, have been increasingly adopted to analyze the structural and dynamical properties of such systems.
Their wider use has increased the need for datasets that can be discovered, interpreted, and reused across domains such as neuroscience, social science, and biology.
Curated repositories therefore form an important part of the research infrastructure: they support empirical analysis, reproducible benchmarking, and comparison across heterogeneous datasets.

Despite these advances, higher-order datasets remain challenging to discover and reuse in a systematic manner, in tension with the broader objective of ensuring that research data are findable, accessible, interoperable, and reusable \citep{Wilkinson:2016}.
In practice, researchers often depend on dispersed sources, personal collections, or labor-intensive reconstruction from raw records.
Differences in file formats, metadata conventions, provenance documentation, and versioning further hinder cross-method comparison and reproducible assessment of dataset suitability.

In this article, we present \texttt{AHORN} (Aachen Higher-Order Repository of Networks), a curated repository for higher-order network research.
\texttt{AHORN} transforms publicly released source datasets into standardized higher-order network artifacts, documents their provenance, and provides access through a web catalog and companion programmatic tooling.

\begin{figure}[t]
    \centering
    \resizebox{\textwidth}{!}{%
\begin{tikzpicture}[
        transform shape,
        font=\footnotesize,
        stage/.style={
            draw=rwth-black-75,
            rounded corners=3pt,
            align=left,
            inner sep=5.5pt,
            line width=0.45pt,
            fill=white,
            text=rwth-black-100
        },
        source-stage/.style={stage, draw=rwth-blue-100, fill=rwth-blue-10},
        process-stage/.style={stage, draw=rwth-green-100, fill=rwth-green-10},
        zenodo-stage/.style={stage, draw=rwth-petrol-100, fill=rwth-petrol-10},
        website-stage/.style={stage, draw=rwth-turquoise-100, fill=rwth-turquoise-10},
        consumer-stage/.style={stage},
        flow/.style={
            -Latex,
            thick,
            draw=rwth-black-75,
            shorten <=1pt,
            shorten >=2pt
        },
        flow-label/.style={
            font=\scriptsize,
            align=center,
            fill=white,
            inner sep=1.5pt,
            text=rwth-black-75
        },
        platform/.style={
            draw=rwth-blue-75,
            dashed,
            rounded corners=4pt,
            inner sep=8pt
        }
    ]
    \node[source-stage, text width=3.5cm] (sources) {
        \textbf{Upstream Datasets}\\
        For example:\\
        $\bullet$ Social data\\
        $\bullet$ Collaboration and review data\\
        $\bullet$ Biomedical data
    };

    \node[process-stage, text width=3.5cm, right=11mm of sources] (contributor) {
        \textbf{Contributor}\\
        $\bullet$ Dataset Converter\\
        $\bullet$ Writes datasheet\\
        $\bullet$ Prepares release files
    };

    \node[process-stage, text width=3.5cm, below=9mm of contributor] (maintainers) {
        \textbf{Maintainers}\\
        $\bullet$ Review submissions\\
        $\bullet$ Validate metadata\\
        $\bullet$ Maintain website
    };

    \node[zenodo-stage, text width=3.5cm, right=11mm of contributor] (zenodo) {
        \textbf{Zenodo repository}\\
        $\bullet$ Versioned data files\\
        $\bullet$ Stable, citable DOI\\
        $\bullet$ Reproducible retrieval
    };

    \node[website-stage, text width=3.5cm, right=11mm of maintainers] (website) {
        \textbf{Browsable website}\\
        $\bullet$ Dataset pages\\
        $\bullet$ Metadata and statistics\\
        $\bullet$ Search and comparison
    };

    \path (zenodo.east) -- (website.east) coordinate[midway] (outcomes-east);

    \node[consumer-stage, text width=3.5cm, right=11mm of outcomes-east] (consumer) {
        \textbf{Consumer / User}\\
        $\bullet$ Browse statistics\\
        $\bullet$ Download and cite data\\
        $\bullet$ Reuse in analyses
    };

    \node[platform, fit=(contributor)(zenodo)(maintainers)(website), label={[font=\bfseries]above:\texttt{AHORN} platform}] {};
    \draw[flow] (contributor.south) -- node[right, flow-label] {submit} (maintainers.north);

    \draw[flow] (sources.east) -- node[above, flow-label] {select} (contributor.west);
    \draw[flow] (contributor.east) -- node[above, flow-label] {upload} (zenodo.west);
    \draw[flow] (maintainers.east) -- node[below, flow-label] {publish} (website.west);
    \draw[flow] (website.north) -- node[right, flow-label] {link} (zenodo.south);
    \draw[flow] (zenodo.east) -- node[pos=0.55, sloped, above, flow-label] {retrieve\\ \& cite} (consumer.north west);
    \draw[flow] (website.east) -- node[pos=0.3, sloped, below, flow-label] {browse} (consumer.south west);
\end{tikzpicture}
    }
    \caption{%
        Overview of the \texttt{AHORN} platform.
        Contributors transform source datasets into a common exchange format and upload versioned files to Zenodo; maintainers review submissions and publish catalog entries that support discovery, citation, programmatic access through \texttt{ahorn-loader}, and downstream reuse.
    }%
    \label{figure:ahorn-overview}
\end{figure}
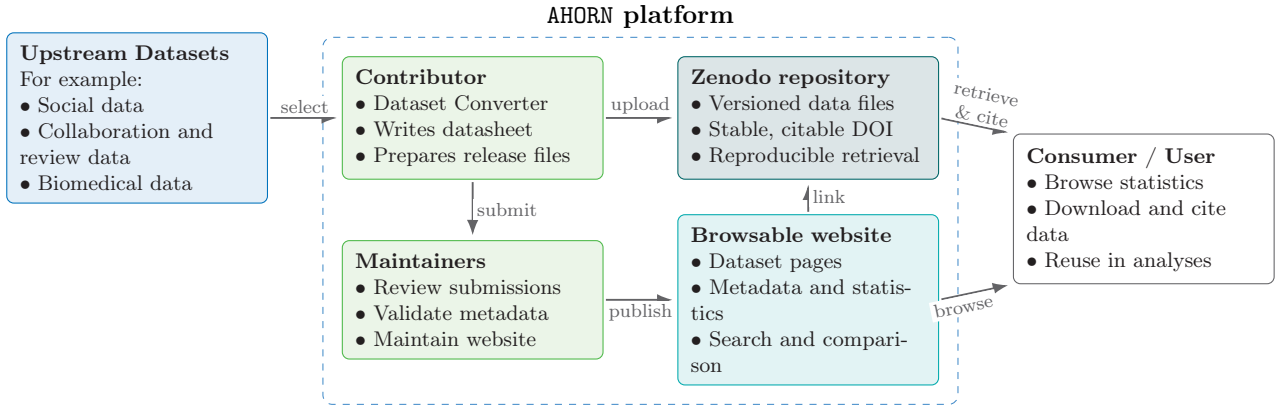

\texttt{AHORN} implements an open-source workflow for dataset submission, validation, and publication.
Its contribution is infrastructural: it provides a repository layer that makes curated higher-order datasets inspectable, citable, and easier to reuse across software ecosystems.
Figure~\ref{figure:ahorn-overview} summarizes \texttt{AHORN}'s role in the dataset lifecycle.

\subsection{Related Work}%
\label{section:related-work}

While substantial repositories for higher-order network data remain scarce, several notable initiatives have emerged in recent years.
Antelmi et al.\ \citep{Dewar:2024} introduced the \texttt{HypergraphRepository}, a web-based collection of hypergraph datasets with basic descriptive statistics and visualization capabilities; however, this resource is currently inaccessible.

Within the \texttt{XGI} software ecosystem \citep{Landry:2023}, the \texttt{XGI-DATA} initiative provides a complementary resource.
Similarly, the recently published \texttt{hypergraphx-data} is maintained as part of the \texttt{hypergraphx} ecosystem \citep{Lotito:2023, Lotito:2026}, offering higher-order datasets designed for integration with the associated library.
However, both initiatives are organized around their respective software stacks rather than constituting independent repository infrastructure.
While \texttt{hypergraphx-data} provides similar discoverability features through a web interface, \texttt{XGI} datasets are accessible only through their software API.

The curated collection maintained by Austin~R.~Benson is another important source of higher-order network datasets \citep{Benson:Data}.
Although not a repository in the strict sense, it has become an essential reference resource for studies of social, biological, and collaboration networks.

Graph repositories are more mature and broader in scope, with established catalogs including \texttt{KONECT}~\citep{Kunegis:2013}, \texttt{SNAP}~\citep{Leskovec:2014}, \texttt{ICON}~\citep{Clauset:2016}, \texttt{Network Repository}~\citep{Rossi:2016}, and \texttt{Netzschleuder}~\citep{Peixoto:2020}.
Although some datasets in these collections admit higher-order interpretations, such structure is only implicit and requires explicit reconstruction by end users.
Furthermore, these repositories do not provide discovery and access mechanisms tailored to higher-order network analysis.
Table~\ref{table:repository-comparison} summarizes these distinctions.

\begin{table}[t]
    \centering
    \caption{%
        Comparison of network repositories with respect to access modality, native support for higher-order data, and software interoperability.
    }
    \label{table:repository-comparison}

\newcommand{\yes}{\ding{51}}
\newcommand{\no}{\ding{55}}
\newcommand{\partialyes}{$\sim$}
\newcommand{\repositorygroup}[1]{
    \begin{tabular}[t]{@{}l@{}}#1
\end{tabular}}

\scriptsize
\setlength{\tabcolsep}{5pt}
\renewcommand{\arraystretch}{1.25}

    \begin{NiceTabular}{p{0.19\linewidth}p{0.19\linewidth}cccccp{0.13\linewidth}}[
        tabularnote={Legend: \yes{} available; \partialyes{} partial or software-specific; \no{} not available.}
    ]
    \toprule
    \Block[l]{2-1}{Repository} & \Block[l]{2-1}{Repository\\model} & \Block{1-2}{Access} & & \Block{2-1}{Dataset\\Stats} & \Block{2-1}{Native\\higher-order} & \Block{2-1}{Availability\\Guarantees\tabularnote{Stable, revision-addressable dataset releases backed by long-term archival hosting rather than mutable repository files.}} & \Block[l]{2-1}{Companion library} \\
    \cmidrule(lr){3-4}
    & & Web & API & & & & \\
    \midrule
    \repositorygroup{\texttt{ICON}, \texttt{SNAP}, \texttt{KONECT},\\\texttt{Network Repository},\\\texttt{Netzschleuder}} &
    Broad network archive &
    \yes & \partialyes & \partialyes & \no & \no & \partialyes \\
    \texttt{HypergraphRepository} &
    Catalog / index &
    \partialyes\tabularnote{The URL reported for \texttt{HypergraphRepository} was unavailable when checked on 29 July 2026.} & \no & \partialyes & \yes & \no & \no \\
    \texttt{XGI-DATA} &
    Library data bundle &
    \partialyes & \yes & \yes & \yes & \yes & \texttt{XGI} \\
    \texttt{hypergraphx-data} &
    Library data bundle &
    \yes & \yes & \yes & \yes & \partialyes & \texttt{hypergraphx} \\
    \texttt{AHORN} &
    Standalone repository &
    \yes & \yes & \yes & \yes & \yes & \texttt{ahorn-loader} \\
    \bottomrule
\end{NiceTabular}

\end{table}

\section{Repository Design}%
\label{section:methods}

\subsection{Curation Workflow}%
\label{section:curation-policy}

\texttt{AHORN} curates datasets derived from publicly released source material rather than collecting new data.
Its scope encompasses resources whose interactions can be represented as simplicial complexes, cell complexes, hypergraphs, or related higher-order network formalisms, and whose redistribution is compatible with the upstream license.
Candidate datasets must have an identifiable upstream source, sufficient metadata for attribution and reuse, and an inspectable and reproducible conversion path.

The existing catalog draws on several source families.
Many entries originate from the higher-order network collection by Benson and collaborators \citep{Benson:Data}, including coauthorship, question-answering, legislative, contact, review, shopping, and recipe datasets.
Additional sources include commonly used citation and coauthorship benchmark files, graph examples available through \texttt{NetworkX} and lifted to higher-order representations, and triangulated manifold data from the \texttt{MANTRA}~\cite{Ballester:2025} benchmark.
This mixture gives the repository both broad domain coverage and a practical test bed for converters that must handle different raw formats, metadata conventions, and interaction semantics.

Every dataset is processed by a source-specific converter script committed with the repository, ensuring reproducibility and traceability of the provided dataset files.
The converters parse the upstream representation, normalize node and interaction identifiers, and write the result in the canonical \texttt{AHORN} exchange format.
For graph sources, converters lift the source graph to the intended higher-order representation, such as a clique complex, while preserving node and edge attributes.
For multi-network or matrix-based sources, preprocessing records additional structure explicitly: the \texttt{MANTRA} converter writes manifold-level network metadata, and the drug-target converter turns interaction and similarity matrices into rank-labeled cells.
The recipe data are published both as a combined cooking dataset and as cuisine-specific child datasets, sharing the same converter and source provenance.

Datasets are incorporated through an open-source workflow in which contributors submit a converter, a dataset page, and the corresponding dataset file.
Loosely inspired by the datasheet framework of Gebru et al.\ \citep{Gebru:2021}, each dataset page records provenance, collection context, recommended citation, licensing or usage constraints, known limitations, and summary statistics.
Maintainers review proposed entries for metadata completeness, converter reproducibility, provenance clarity, validation success, and consistency with repository conventions.
Accepted datasets are released as versioned artifacts, with prior releases retained to support stable citation and reproducible analyses in line with established data-citation practices \citep{Martone:2014}.
If errors, upstream changes, or rights-holder objections are reported, maintainers can revise metadata, publish a corrected revision, or remove redistributed files from public access.

\subsection{Access and Data Format}%
\label{section:access-and-format}

\texttt{AHORN} exposes repository content through two access routes.
The public website provides a browsable catalog with search and filter functions for human discovery.
Catalog entries provide source information, provenance, citation guidance, downloadable files, and visual summaries where supported by the metadata.
Published revisions are archived as versioned Zenodo records, providing persistent landing pages and revision-addressable access.

In parallel with the human-readable web interface, \texttt{AHORN} publishes a machine-readable dataset index at \url{https://ahorn.rwth-aachen.de/api/datasets.json}.
This endpoint exposes dataset identifiers, titles, tags, and files for programmatic discovery and retrieval.

\texttt{AHORN} separates citation of the repository infrastructure from citation of individual dataset listings.
Users who refer to \texttt{AHORN} as a repository or platform should cite this article.
Users who analyze a specific dataset should cite the original source publication listed on the \texttt{AHORN} dataset page, possibly together with the corresponding Zenodo record so that the exact dataset version can be identified.

Datasets are distributed in a line-based plain-text format, currently version 0.3, with optional gzip compression.
The first line contains dataset-level metadata as a JSON object.
Subsequent lines define explicit nodes or interactions; an interaction is represented by a comma-separated list of node identifiers, and either record type may be followed by a JSON attribute object.
Nodes without attributes can be declared implicitly through their interactions.
Multi-network files use JSON marker lines to delimit constituent networks.
Standardized attributes include weights and ISO~8601 timestamps, while dataset-specific attributes can represent labels or other source information.
The explicit versioning of the format allows the specification to evolve in a controlled manner.

Compared with related interchange efforts such as the Hypergraph Interchange Format \citep{Coll:2025}, the canonical \texttt{AHORN} format prioritizes low-friction publication and human readability.
For datasets whose structure is covered by the HIF schema, \texttt{AHORN} additionally publishes HIF exports, enabling reuse by HIF-compatible software such as \texttt{XGI} \citep{Landry:2023}, \texttt{HyperNetX} \citep{Praggastis:2024}, and \texttt{SimpleHypergraphs.jl} \citep{Spagnuolo:2020}.
The full specification, including examples and required fields, is maintained at \url{https://ahorn.rwth-aachen.de/about/format/}.

\section{Access and Validation}%
\label{section:ahorn-loader}

To facilitate the integration of \texttt{AHORN} into reproducible research workflows, we developed \texttt{ahorn-loader}, which is provided both as a Python library and as a language-agnostic command-line interface for programmatic repository access.
The tool supports two tasks:
\begin{enumerate*}[label=(\roman*)]
\item retrieval of (revision-pinned) datasets and
\item pre-submission validation of dataset files against \texttt{AHORN} requirements.
\end{enumerate*}

\texttt{ahorn-loader} provides a lightweight interface for browsing the catalog and downloading datasets.
Downloads are cached locally, and datasets in the canonical \texttt{AHORN} format can be opened as iterable text streams for direct use in analysis scripts.
Where available, other dataset formats can be downloaded through an explicit format switch.
Revision-pinned calls allow benchmark workflows to retrieve fixed dataset snapshots without new dataset revisions affecting reproducibility.

The built-in validator ensures that datasets adhere to the specification (cf. Section~\ref{section:access-and-format}) and running it is a prerequisite for adding any new dataset to the repository.
It checks syntactic validity, proper use of standardized metadata fields, and structural compliance with the dataset format.
Automated tests cover both the access API and validator.

\section{Repository Coverage}%
\label{section:repository-content}

At the time of writing this article, \texttt{AHORN} contains \num{83} datasets with \num{94} revisions.
The catalog spans four formalism groups, multiple application domains, and both static and temporal settings, including social contact networks, collaboration and citation data, consumer-review systems, biomedical resources, and geometric benchmark collections.
This coverage makes \texttt{AHORN} useful both as a source of individual benchmark files and as a cross-domain corpus.

For each dataset, we measured the number of unique nodes, the number of interactions, the interaction-size profile, the mean and maximum interaction size, the share of interactions of size greater than two, the share of unique interaction member sets, as well as the share of active nodes that participate in at least one interaction, the interaction-membership density per node, and the effective number of interaction sizes derived from the entropy of the interaction-size distribution.

Repository composition is heterogeneous but centered on datasets that support straightforward higher-order interpretations.
Of the \num{83} datasets, \num{75} are annotated as hypergraphs, \num{44} are simplicial complexes, and one is a combinatorial complex (multi-selection possible).
Temporal structure is present in \num{15} datasets.
The groups also differ in structural scale: the median maximum interaction size is \num{5.0} for simplicial complexes, \num{30.0} for hypergraphs, and \num{81.0} for datasets with multiple types; the single combinatorial complex reaches \num{315}.

Moreover, metadata coverage is high for the core discovery fields.
Figure~\ref{figure:domain-metadata-distribution} summarizes the distribution of application domains and the presence of metadata across datasets.

\begin{figure}[t]
    \centering
\pgfplotstableread[col sep=comma]{generated/domain_distribution.csv}\domaindistribution
\pgfplotstableread[col sep=comma]{generated/metadata_distribution.csv}\metadatadistribution
\begin{tikzpicture}
    \begin{groupplot}[
            group style={
                group size=2 by 1,
                horizontal sep=1.4cm,
            },
            width=0.5\linewidth,
            height=0.35\linewidth,
            ymin=0,
            xtick=data,
            xtick align=outside,
            x tick label style={font=\scriptsize},
            yticklabel style={font=\footnotesize},
            enlarge x limits=0.12,
        ]
        \nextgroupplot[
            title={Domain Distribution},
            ybar stacked,
            xticklabels from table={\domaindistribution}{domain},
            ylabel={Count},
            legend to name=network-type-legend,
            legend style={
                font=\tiny,
                legend columns=4,
                draw=none,
                /tikz/every even column/.append style={column sep=0.5em},
            },
            legend cell align=left,
            cycle list name=rwth,
            x tick label style={rotate=25, anchor=east, font=\scriptsize},
        ]
        \addplot table[
            x expr=\coordindex,
            y={hypergraph},
        ] {\domaindistribution};
        \addlegendentry{Hypergraph}

        \addplot table[
            x expr=\coordindex,
            y={simplicial_complex},
        ] {\domaindistribution};
        \addlegendentry{Simplicial Complex}

        \addplot table[
            x expr=\coordindex,
            y={combinatorial_complex},
        ] {\domaindistribution};
        \addlegendentry{Combinatorial Complex}

        \addplot table[
            x expr=\coordindex,
            y={multiple},
        ] {\domaindistribution};
        \addlegendentry{Multiple}

        \nextgroupplot[
            title={Metadata Distribution},
            ybar stacked,
            xticklabels from table={\metadatadistribution}{metadata},
            cycle list name=rwth,
        ]
        \addplot table[
            x expr=\coordindex,
            y={hypergraph},
        ] {\metadatadistribution};

        \addplot table[
            x expr=\coordindex,
            y={simplicial_complex},
        ] {\metadatadistribution};

        \addplot table[
            x expr=\coordindex,
            y={combinatorial_complex},
        ] {\metadatadistribution};

        \addplot table[
            x expr=\coordindex,
            y={multiple},
        ] {\metadatadistribution};
    \end{groupplot}
    \node[anchor=north] at ($(group c1r1.south)!0.5!(group c2r1.south)-(0,1.0cm)$) {\pgfplotslegendfromname{network-type-legend}};
\end{tikzpicture}
    \caption{%
        Catalog-level distributions across datasets, stratified by network type.
        The left panel depicts the distribution across application domains, whereas the right panel summarizes metadata availability.
    }%
    \label{figure:domain-metadata-distribution}
\end{figure}
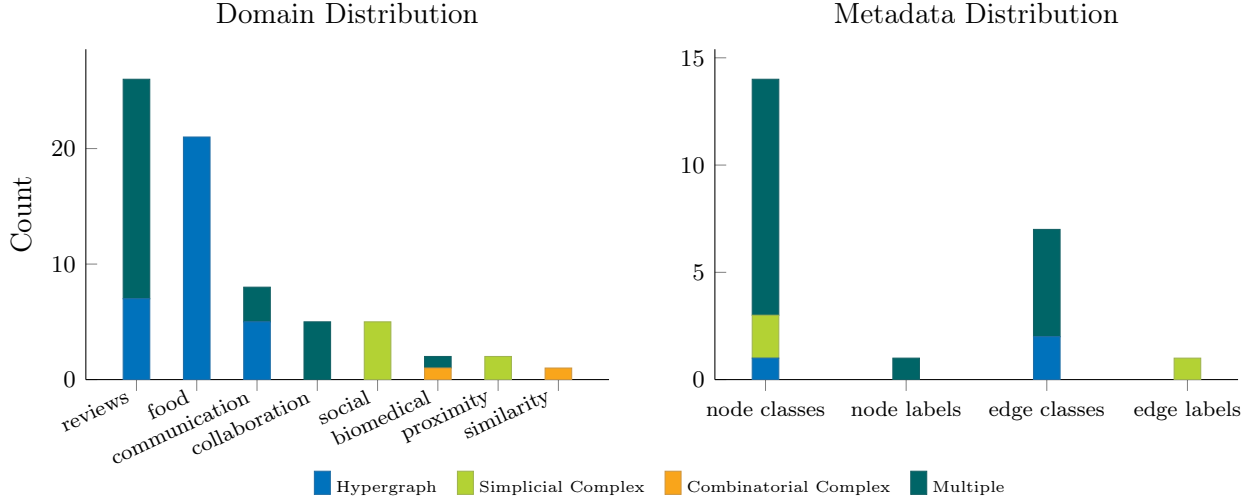

\begin{figure}[t]
    \centering
    \pgfplotstableread[col sep=comma]{generated/dataset_scale_scatter.csv}\datasetscaledata
\pgfplotstableread[col sep=comma]{generated/interaction_size_frequency.csv}\interactionsizefrequencydata
\pgfplotsset{
    ahorn scatter point/.style={
        only marks,
        mark size=1.9pt,
        forget plot,
    },
    ahorn hypergraph/.style={
        mark options={fill=rwth-blue-100,draw=black},
    },
    ahorn simplicial/.style={
        mark options={fill=rwth-maygreen-100, draw=black},
    },
    ahorn combinatorial/.style={
        mark options={fill=rwth-orange-100,draw=black},
    },
    ahorn multiple/.style={
        mark options={fill=rwth-petrol-100,draw=black},
    },
    ahorn static/.style={mark=*},
    ahorn temporal/.style={mark=triangle*},
}
\begin{tikzpicture}
    \begin{groupplot}[
            group style={group size=2 by 1, horizontal sep=1.75cm},
            width=0.46\linewidth,
            height=0.38\linewidth,
            tick label style={font=\footnotesize},
            label style={font=\footnotesize},
        ]
        \nextgroupplot[
            title={Dataset scale},
            xmode=log,
            ymode=log,
            xlabel={Number of Nodes},
            ylabel={Number of Interactions},
            legend style={font=\scriptsize, at={(0.5,-0.28)}, anchor=north, row sep=1pt, column sep=0.8em, draw=none, fill=none},
            legend columns=2,
            legend cell align=left,
        ]
        \addplot[ahorn scatter point, ahorn simplicial, ahorn static] table[x=simplicial_static_nodes, y=simplicial_static_interactions] {\datasetscaledata};
        \addplot[ahorn scatter point, ahorn simplicial, ahorn temporal] table[x=simplicial_temporal_nodes, y=simplicial_temporal_interactions] {\datasetscaledata};
        \addplot[ahorn scatter point, ahorn hypergraph, ahorn static] table[x=hypergraph_static_nodes, y=hypergraph_static_interactions] {\datasetscaledata};
        \addplot[ahorn scatter point, ahorn hypergraph, ahorn temporal] table[x=hypergraph_temporal_nodes, y=hypergraph_temporal_interactions] {\datasetscaledata};
        \addplot[ahorn scatter point, ahorn combinatorial, ahorn static] table[x=combinatorial_static_nodes, y=combinatorial_static_interactions] {\datasetscaledata};
        \addplot[ahorn scatter point, ahorn combinatorial, ahorn temporal] table[x=combinatorial_temporal_nodes, y=combinatorial_temporal_interactions] {\datasetscaledata};
        \addplot[ahorn scatter point, ahorn multiple, ahorn static] table[x=multiple_static_nodes, y=multiple_static_interactions] {\datasetscaledata};
        \addplot[ahorn scatter point, ahorn multiple, ahorn temporal] table[x=multiple_temporal_nodes, y=multiple_temporal_interactions] {\datasetscaledata};

        \addlegendimage{legend image code/.code={}}
        \addlegendentry{\textbf{Network type}}
        \addlegendimage{legend image code/.code={}}
        \addlegendentry{\textbf{Temporal status}}

        \addlegendimage{only marks, mark=*, ahorn hypergraph}
        \addlegendentry{Hypergraph}

        \addlegendimage{only marks, ahorn static, mark options={fill=black, draw=black, line width=0.35pt}}
        \addlegendentry{static}

        \addlegendimage{only marks, mark=*, ahorn simplicial}
        \addlegendentry{Simplicial Complex}

        \addlegendimage{only marks, ahorn temporal, mark options={fill=black, draw=black, line width=0.35pt}}
        \addlegendentry{temporal}

        \addlegendimage{only marks, mark=*, ahorn combinatorial}
        \addlegendentry{Combinatorial Complex}

        \addlegendimage{legend image code/.code={}}
        \addlegendentry{}

        \addlegendimage{only marks, mark=*, ahorn multiple}
        \addlegendentry{Multiple}

        \addlegendimage{legend image code/.code={}}
        \addlegendentry{}

        \nextgroupplot[
            title={Interaction-size distribution},
            xmin=1,
            xmax=20.5,
            ymin=0,
            xlabel={Interaction size},
            ylabel={Median share of interactions (\%)},
        ]
        \addplot[draw=rwth-blue-100, mark=*, mark size=1.6pt, mark options={fill=rwth-blue-50, draw=rwth-blue-100}] table[x=interaction_size, y=median_share_pct] {\interactionsizefrequencydata};
    \end{groupplot}
\end{tikzpicture}
    \caption{%
        Dataset scale and interaction-size distributions across the repository.
        The \emph{Dataset scale} panel presents individual datasets, stratified by their number of nodes and interactions, with coloring denoting formalism groups and marker shapes differentiating between temporal and static datasets.
        The \emph{Interaction-size distribution} panel illustrates the median per-dataset proportion of interactions as a function of interaction size (truncated at $20$).
    }%
    \label{figure:dataset-scale-scatter}
\end{figure}
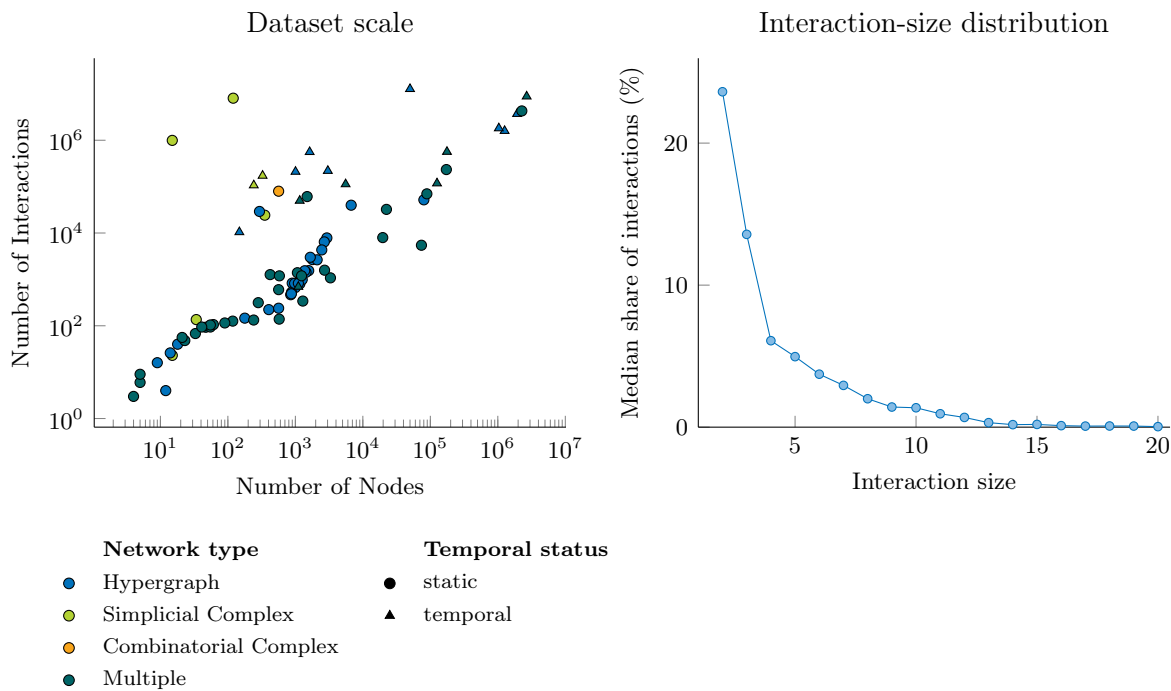

Scale varies over several orders of magnitude (Figure~\ref{figure:dataset-scale-scatter}).
The datasets range from \num{4} to \num{2321767} nodes and from \num{3} to \num{12741779} interactions.
This breadth reflects both compact benchmarks for method prototyping and larger empirical case studies.
The scatter also highlights structurally distinct outliers, including MANTRA collections with many simplices despite small per-instance node sets and large coauthorship or review datasets with millions of nodes and interactions.
Typical datasets are nevertheless dominated by small interactions: the median per-dataset share is \num{23.6}\% for size \num{2} and \num{13.6}\% for size \num{3}, and generally decreases for larger exact interaction sizes.

Additional heterogeneity emerges beyond raw scale.
Across the full catalog, the median share of unique interactions is \num{98.6}\%, while the median active-node share is \num{100.0}\%.
These medians hide several distinct reuse regimes.
The MANTRA collections and the temporal contact datasets are dominated by repeated member sets, with unique-interaction shares between \num{0.04}\% and \num{11.9}\%, whereas benchmarks such as MAG-10 and Drug-Target Interaction contain only unique interaction sets.
Likewise, several citation-style datasets contain many declared but inactive nodes in the released file, including PubMed Co-Citation with an active-node share of \num{19.5}\% and CiteSeer Co-Citation with \num{44.0}\%.
The effective number of interaction sizes ranges from \num{1.0} in the fixed-size MANTRA collections to \num{5.0} in Vegas Bars Reviews (Dance Clubs), showing that the current repository spans both tightly constrained and broadly distributed higher-order interaction patterns.

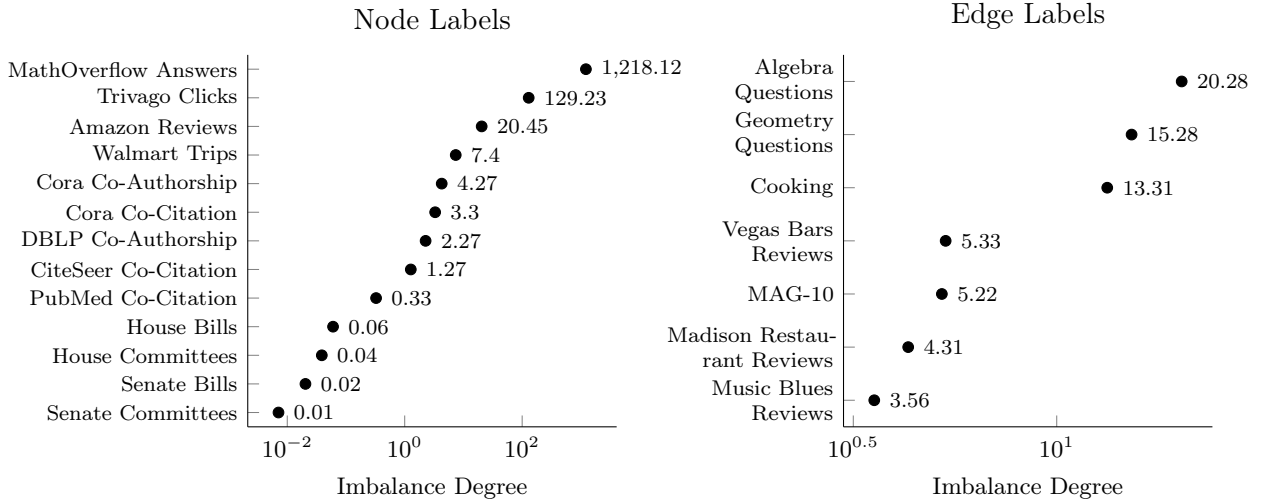
\begin{figure}[t]
    \centering

\pgfplotstableread[col sep=comma]{generated/label_imbalance.csv}\labelimbalancedata

\newif\iflabelimbalancefirst
\newcommand{\collectlabelimbalancelabels}[2]{%
    \def#2{}%
    \labelimbalancefirsttrue
    \pgfplotstableforeachcolumnelement{metadata}\of\labelimbalancedata\as\metadatavalue{%
        \edef\labelimbalancecurrent{\metadatavalue}%
        \edef\labelimbalancetarget{#1}%
        \ifx\labelimbalancecurrent\labelimbalancetarget
        \pgfplotstablegetelem{\pgfplotstablerow}{dataset}\of\labelimbalancedata%
        \iflabelimbalancefirst
        \xdef#2{{\pgfplotsretval}}%
        \global\labelimbalancefirstfalse
        \else
        \xdef#2{#2,{\pgfplotsretval}}%
        \fi
        \fi
    }%
}

\collectlabelimbalancelabels{node}{\nodelabelimbalanceyticklabels}
\collectlabelimbalancelabels{edge}{\edgelabelimbalanceyticklabels}

\begin{tikzpicture}
    \begin{groupplot}[
            group style={
                group size=2 by 1,
                horizontal sep=3cm,
            },
            width=0.38\linewidth,
            height=6.5cm,
            xmode=log,
            xlabel={Imbalance Degree},
            ytick=data,
            y dir=reverse,
            enlarge y limits={abs=0.5},
            tick label style={font=\footnotesize},
            label style={font=\footnotesize},
            y tick label style={font=\scriptsize},
            nodes near coords={
                \pgfmathprintnumber[fixed, precision=2]{\pgfplotspointmeta}
            },
            every node near coord/.append style={
                font=\scriptsize,
                anchor=west,
                xshift=2pt,
            },
            point meta=explicit,
        ]
        \nextgroupplot[
            title={Node Labels},
            yticklabels/.expanded={\nodelabelimbalanceyticklabels},
        ]
        \addplot[
            only marks,
            mark=*,
        ] table[
            col sep=comma,
            discard if not={metadata}{node},
            meta=imbalance,
            x=imbalance,
            y=index,
        ] {generated/label_imbalance.csv};

        \nextgroupplot[
            title={Edge Labels},
            yticklabels/.expanded={\edgelabelimbalanceyticklabels},
            y tick label style={font=\scriptsize, text width=2.3cm, align=right},
        ]
        \addplot[
            only marks,
            mark=*,
        ] table[
            col sep=comma,
            discard if not={metadata}{edge},
            meta=imbalance,
            x=imbalance,
            y=index,
        ] {generated/label_imbalance.csv};
    \end{groupplot}
\end{tikzpicture}
    \caption{%
        Label imbalance degrees (using Euclidean distance) for all datasets with node or edge labels.
    }%
    \label{figure:label-imbalance}
\end{figure}

For datasets with node or edge labels, label imbalance \citep{OrtigosaHernandez:2017} is reported in Figure~\ref{figure:label-imbalance}.
Within that subset, node-labeled datasets span a wide range from \num{0.007} in Senate Committees to \num{1218.1} in MathOverflow Answers, whereas edge-labeled datasets occupy a narrower band between \num{3.56} and \num{20.28}.
This view complements the structural profile by showing how concentrated the class distributions are in benchmark-style supervised datasets.

The complete dataset catalog is provided as supplementary material.

\section{Discussion}%
\label{section:discussion}

\texttt{AHORN} provides a curated repository layer for higher-order network data that combines standardized file formats with provenance records, citable and stable versioning, and easy access routes.
Together, these components address a practical gap in higher-order network research, where datasets are often reused without a shared mechanism for discovery, revision tracking, or inspection of the derived representation.

The repository overview shows that \texttt{AHORN} already spans several distinct settings: compact pedagogical benchmarks, repeated-event temporal datasets, and large collections with millions of nodes or interactions.
Because entries are linked to upstream sources and released as stable revisions, they can be compared reproducibly across workflows and libraries.

Several limitations remain.
Domain coverage is shaped by the availability of public sources, converter effort, documentation quality, and redistribution permissions.
Metadata completeness also depends partly on upstream documentation, which is reflected in uneven coverage of citation and licensing information.
The catalog summaries treat parent datasets and their derived sub-datasets as separate observations; they therefore describe the repository's published inventory rather than a set of statistically independent source datasets.
The current exchange-format specification does not yet standardize directed interactions or trajectory data.
Finally, a dataset page documents the known upstream license or usage conditions but does not grant rights beyond those conditions.

\texttt{AHORN} is already integrated into \texttt{TopoNetX} \citep{Hajij:2024}, allowing users of that library to access and analyze higher-order datasets through familiar APIs.
The publication of HIF exports provides a complementary interoperability route for HIF-compatible libraries and workflows.
Further interoperability with community software and continued expansion of metadata and license coverage should improve reuse while preserving the repository's focus on transparent curation and stable access.

\section{Data and Code Availability}

The \texttt{AHORN} web platform is accessible at \url{https://ahorn.rwth-aachen.de/}, with machine-readable discovery metadata available at \url{https://ahorn.rwth-aachen.de/api/datasets.json}.
Dataset pages link to versioned files hosted via Zenodo (\url{https://zenodo.org/communities/ahorn/}), preserving individual releases as stable snapshots.

The source code for the platform and dataset processing is publicly available at \url{https://github.com/netsci-rwth/ahorn} under the GNU Affero General Public License v3.0.
The companion Python package \texttt{ahorn-loader}, which provides download and validation functions, is available at \url{https://github.com/netsci-rwth/ahorn-loader} under the MIT License and is distributed via PyPI (\url{https://pypi.org/project/ahorn-loader/}).

\bibliographystyle{naturemag}
\bibliography{references}




\section*{Funding}

The authors acknowledge funding by the European Union (ERC, HIGH-HOPeS, 101039827).
Views and opinions expressed are however those of the authors only and do not necessarily reflect those of the European Union or the European Research Council Executive Agency.
Neither the European Union nor the granting authority can be held responsible for them.

\section*{Ethics Statement}

This study did not generate new human or animal data.
Ethical approval, informed consent, anonymization, and study-specific governance remain with the original data creators and are described in the corresponding source publications or repository records where applicable.
Inclusion in \texttt{AHORN} does not replace the ethical and legal assessment required for a specific downstream use.

\end{document}